\documentclass[twocolumn,pra,amssymb,showpacs]{revtex4-1}
\usepackage{graphicx}
\usepackage{amsmath}
\usepackage{subfigure}

\newcommand{\be}{\begin{equation}}
\newcommand{\ee}{\end{equation}}
\newcommand{\ben}{\begin{eqnarray}}
\newcommand{\een}{\end{eqnarray}}

\begin{document}

\title{Solving the two dimensional Schr\"odinger equation using basis truncation: a controversial case}

\author{J. Batle$^1$}
\email{E-mail address: jbv276@uib.es}
\affiliation{ 
$^1$Departament de F\'{\i}sica, Universitat de les Illes Balears, 07122 Palma de Mallorca, Balearic Islands, Spain \\\\}

\date{\today}

\begin{abstract}

Solutions of the Schr\"odinger equation by spanning the wave function is a complete basis is a common practice is many-body interacting systems. We shall study the case of a two-dimensional quantum system composed by two interacting spin-less electrons and see that the correctness of the matrix approach depends inexplicably on the type of interaction existing between particles.

\end{abstract}


\maketitle

\section{Introduction}

A common textbook quantum physics/chemistry approach to solving the Schr\"odinger equation in those cases where no analytical solution is available is to utilize matrix mechanics \cite{llibreCristian}. In atomic and molecular electronic structure calculations, it is often important to go beyond the independent-particle approximation, and thus some numerical machinery is required. 
Vast numbers of quantum mechanical research problems that
are amenable to solution have been solved using matrix mechanics, one good example being for instance the study of electrons
on a small lattice \cite{Dagotto}. One decomposes the wave function into a complete set 
of well known basis states

\begin{equation}
|\psi\rangle=\sum^\infty_{i=1}a_i|\psi_i\rangle,
\end{equation}

\noindent where the $a_i$'s are the unknown coefficients. Inserting this into the time-independent Schr\"odinger equation, and undergoing inner products with the same basis states yields the eigenvalue equation

\begin{equation}
\sum_{j=1}^\infty H_{ij} a_j = E a_i,
\end{equation}

\noindent where the matrix elements are given by

\begin{equation}
H_{ij} = \langle \psi_i|H|\psi_j \rangle.
\label{HM}   
\end{equation}

\noindent Usually, one just simply truncates the expansion to include only the 
low-lying bound states. If the basis is reasonably chosen, a few states are required to provide satisfactory results. We have to mention that some studies \cite{llibreCristian} have shown that basis set truncation error is of more importance than truncation of the corresponding perturbation series. However, we will not be dealing with such situation here.\noindent

In the present contribution we shall tackle the unexplained behavior of the matrix formalism depending on the Hamiltonian of a problem. In Section II we introduce a model Hamiltonian in two dimensions. In Section III we present the numerical 
approach reached where the matrix formalism works perfectly well. The introduction of a brand new system possessing analytical solution is done in Section IV, which constitutes a terrible unexplained failure of the method of basis expansion. Finally, some conclusions are drawn in Section V.

\section{The model}

We shall provide here a simple system of two interacting spin-less electrons. 
Let us suppose that we have two concentric rings. Electron 1 is located on the inner ring of radius $R_1$ and electron 2 is located on the outer ring of radius $R_2$. Positions are determined by $\phi_1$ and $\phi_2$, respectively. We shall assume $R_1 \leq R_2$. The distance between them is given by $d(\phi_1,\phi_2)=\sqrt{R_1^2+R_2^2-2R_1R_2\cos(\phi_1-\phi_2)}$. 

The corresponding Schr\"odinger equation with electrons interacting via Coulomb potential reads as

\begin{eqnarray} \label{equation}
&&-\frac{1}{2R_1^2}\frac{\partial^2}{\partial\phi_1^2} \Psi(\phi_1,\phi_2) -\frac{1}{2R_2^2}\frac{\partial^2}{\partial\phi_2^2} \Psi(\phi_1,\phi_2) \cr
&&+ \frac{1}{d(\phi_1,\phi_2)} \Psi(\phi_1,\phi_2) \,=\, E  \Psi(\phi_1,\phi_2), 
\end{eqnarray}

\noindent where $\phi_i \in [0,2\pi)$. The solution is obviously periodic $\Psi(0,\phi_2)\,=\,\Psi(2\pi,\phi_2)$, $\Psi(\phi_1,0)\,=\,\Psi(\phi_1,2\pi)$, with $R_1R_2\int_0^{2\pi} \int_0^{2\pi} d\phi_1d\phi_2 \, | \Psi(\phi_1,\phi_2)|^2\,=\,1$. This case is quasi-exactly solvable by using the distance between particles as a new variable \cite{Pierre}. We shall test our numerical procedure with one exact case in order to check the validity of our approach to the problem.

\section{Numerical approach}

One easy way of preserving periodicity is to span the solution $\Psi(\phi_1,\phi_2)$ in the basis of non-interacting two particles, one in each ring, and then truncate the expansion to $N+1$ terms, $N$ even. That is,

\begin{equation} \label{spanned}
\Psi(\phi_1,\phi_2)\,=\,\sum_{m=-\frac{N}{2}}^{\frac{N}{2}} \sum_{n=-\frac{N}{2}}^{\frac{N}{2}} c_{m,n}\,\frac{1}{2\pi\sqrt{R_1 R_2}} e^{im\phi_1}e^{in\phi_2}
\end{equation}

\noindent Had we considered concentric spheres, we should be dealing with spherical harmonics. Plugging (\ref{spanned}) into (\ref{equation}), multiplying by $\frac{1}{2\pi\sqrt{R_1 R_2}} e^{-ik\phi_1}e^{-il\phi_2}$ and integrating over $\{\phi_1,\phi_2\}$ returns

\begin{eqnarray} \label{set}
\sum_{k=-\frac{N}{2}}^{\frac{N}{2}} \sum_{l=-\frac{N}{2}}^{\frac{N}{2}}  && \bigg[\bigg(\frac{m^2}{2R_1^2} + \frac{n^2}{2R_2^2} \bigg)\delta_{k,m}\delta_{l,n} + \langle kl | \frac{1}{d} |mn\rangle \cr
&&-E\,\delta_{k,m}\delta_{l,n} \bigg]\,c_{k,l}\,=\,0,
\end{eqnarray}

\noindent for $m,n\,=\,-\frac{N}{2},..,\frac{N}{2}$. Let us regard $H_{klmn}$ the first line in (\ref{set}). Solving (\ref{set}) for $c_{k,l}$ is tantamount as providing an approximate 
solution to (\ref{equation}) for the ground or excited states, increasing the accuracy when augmenting the number of terms in the expansion $N+1$.

\noindent The matrix element in (\ref{set}) reads explicitly as

\begin{equation}
 \langle kl | \frac{1}{d} |mn\rangle\,=\,\frac{1}{4\pi^2} \int_0^{2\pi} \int_0^{2\pi} d\phi_1d\phi_2 \,\frac{e^{i(m-k)\phi_1}e^{i(n-l)\phi_2}}{d(\phi_1,\phi_2)}.
\end{equation}

\noindent The set of Eqs. (\ref{set}) for $c_{k,l}$ does not read yet as a standard eigenvalue problem. Usual approaches to matrix quantum mechanics deal with only one quantum number, either because the instances addressed are one-dimensional problems or physical scenarios with higher spatial dimensions but characterized with only one principal quantum number. We have to point out that when this is not the case, not a single textbook explains, to our knowledge, how to proceed. 

 In order to tackle the problem given by (\ref{set}), we shall transform 
$H_{klmn} \longrightarrow A_{ij}$ and $c_{k,l} \longrightarrow g_j$, $i,j\,=\,1,..,(N+1)^2$ using $i=(m+\frac{N}{2})(N+1)+(n+\frac{N}{2})+1$ and $j=(k+\frac{N}{2})(N+1)+(l+\frac{N}{2})+1$ $\forall\,(k,l,m,n)$. Notice that by doing so, the problem increases significantly the effective total dimension of the ensuing eigenvalue problem. Also, it is straightforward to extend the previous linear mapping of indexes to more quantum numbers if required. However, if that was the case, the final computational problem becomes quite involved.

With the previous transformation, we have the usual eigenvalue and eigenvector problem

\begin{equation}
\sum_{j=1}^{(N+1)^2} \big( A_{ij} - E\,\delta_{ij} \big)\,g_j\,=\,0,
\end{equation}

\noindent and $i\,=\,1,2,..,(N+1)^2$. Finding the corresponding eigenvalues will give as the energy spectrum of the system. In order to find the eigenvectors, the inverse transformation $g_j \longrightarrow c_{k,l}$ can be proved to be unique. In other words, given $j$ and $N$, we find a sole couple ($k,l$). In practice, we have to solve a linear diophantine equation.\newline

In order to validate our numerical results, we can compare with the analytic case of two concentric rings \cite{Pierre}. Results are shown in Table I. The matching is perfect. 

\begin{table}[h]
\begin{center}
\begin{tabular}{|c||c|c|}
  \hline
  $k$ & $l$ & $c_{k,l}$ \\
  \hline
  -5 & 5 &  4.52937008E-005 \\
   -4 & 4 & 0.000210684568  \\
  -3 & 3 &  0.00133573123  \\
  -2 & 2 &  0.0393656555 \\
  -1 & 1 & -0.401700424  \\
   0 & 0 &  0.821078904  \\
   1 & -1 & -0.401700424 \\
   2 & -2 & 0.0393656555 \\
   3 & -3 & 0.00133573123 \\
   4 & -4 & 0.000210684568 \\
   5 & -5 & 4.52937008E-005 \\
   \hline
\end{tabular}
\end{center}
\caption{Solution coefficients $c_{k,l}$ for the analytic case $R_1 = \frac{13}{7} \sqrt{3 (13 - \sqrt{78})}$, 
$R_2 = \frac{13}{7} \sqrt{3 (13 + \sqrt{78})}$, $\alpha=0$ and $H=0$. Numerical ground energy is virtually ${\it equal}$ to exact 
energy $\frac{28}{507}$. Notice the symmetry in the 
indexes $k,l$ and in the numerical value of $c_{k,l}$. As we can appreciate, only 11 coefficients $c_{k,l}$ 
in the expansion suffice to find the right solution. See text for details.}
\end{table}

The symmetry in the coefficients has a two-fold meaning: on the one hand, the total truncated state is real, whereas on the other hand, the system depends only on the difference of angles $|\phi_1-\phi_2|$.

The method of spanning the function in a suitable basis proves to be very much convenient. Although numerical, it becomes an exact eigenvalue problem when the number of truncated elements $N$ tends to infinity.

\section{An analytical (and pathological) counterexample}

Let us suppose now that our system is not interacting via Coulomb repulsion, but under the action of a harmonic potential between particles. 
The corresponding Schr\"odinger equation to solve is thus given by

\begin{eqnarray} \label{equationw}
&&-\frac{1}{2R_1^2}\frac{\partial^2}{\partial\phi_1^2} \Psi(\phi_1,\phi_2) -\frac{1}{2R_2^2}\frac{\partial^2}{\partial\phi_2^2} \Psi(\phi_1,\phi_2) \cr
&&+ \frac{1}{2}\Omega^2 \big[R_1^2+R_2^2-2R_1R_2\cos(\phi_1-\phi_2)\big] \Psi(\phi_1,\phi_2) \cr
&&\,=\, E  \Psi(\phi_1,\phi_2).
\end{eqnarray}

Introducing $\omega=\phi_1-\phi_2$, we obtain 

\begin{equation}\label{equation2}
\frac{d^2}{d\omega^2} \Psi(\omega) \,+\,A \cos\omega\, \Psi(\omega)\,+\,B \Psi(\omega)\,=\,0,
\end{equation}
\noindent with $\omega \in [0,\pi]$. Defining $\frac{1}{\sigma^2} \equiv \frac{1}{R_1^2}+\frac{1}{R_2^2}$, we have 
$A=2R_1R_2\Omega^2\sigma^2$ and $B=2\sigma^2\big[ E-\frac{1}{2}\Omega^2(R_1^2+R_2^2) \big]$

The solution to ({\ref{equation2}}) is analytic, and given by

\begin{equation}
\Psi(\omega)\,=\,S\bigg(4B,-2A,\frac{\omega}{2}\bigg),
\end{equation}

\noindent where $S$ is the sine elliptic odd Mathieu function. For nonzero $-2A$, the Mathieu functions are only periodic in $\omega$ for certain 
values of $4B$, and this is how the energy is quantized. Such characteristic values are expressed as $b_{2(n-1)}$, $n$ being a natural number (actually it is the number of nodes in the wave function between 0 and $\pi$). The values of $b_{2(n-1)}$ depend on 
$(-2A=)\,- 4R_1R_2\Omega^2\sigma^2$. The final quantized energies for (\ref{equation}) read as

\begin{equation}
E_n\,=\,\frac{1}{2}\Omega^2(R_1^2+R_2^2) \,+\, \frac{b_{2(n-1)}}{8\sigma^2}, \,\,\,n=0,1,2..
\end{equation}

This exact system has not been considered in the past, and reduces to the case studied in \cite{Pierre2} for $R_1=R_2=R$.\newline

For the sake of comparison, let us assume $R_1=1$, $R_2=2$ and $\Omega=1$. This makes $\sigma^2=\frac{4}{5}$, 
$ - 4R_1R_2\Omega^2\sigma^2=-\frac{32}{5}$ and $E_n=\frac{5}{2}+\frac{5}{32} b_{2(n-1)}$. 
Since $b_{-2}\big( -\frac{32}{5}\big)=1.0274$, the ground state energy becomes $E_0=2.660$ (a.u.).

\begin{figure}[htbp]
\begin{center}
\includegraphics[width=8.8cm]{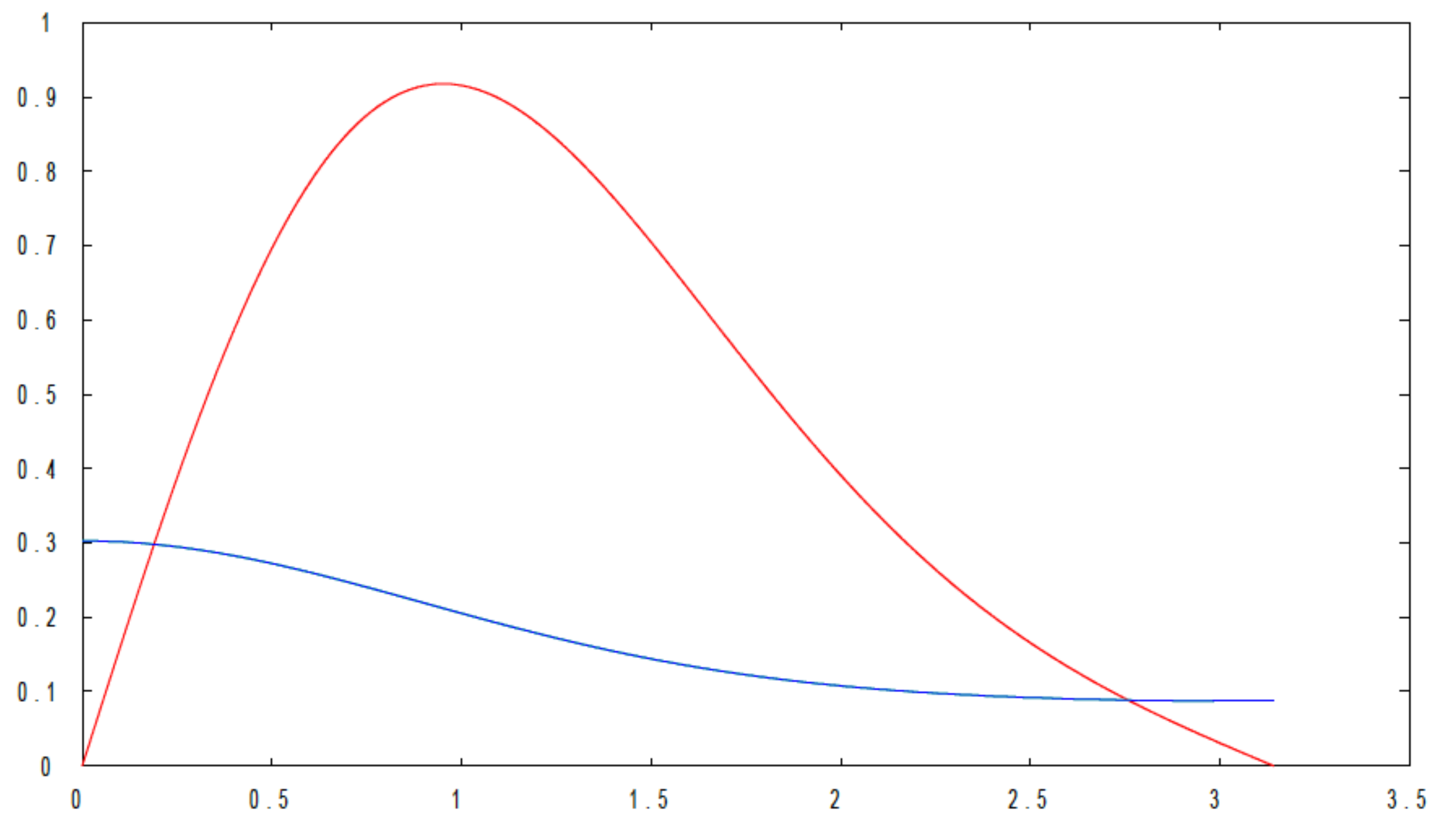}
\caption{(Color online) Exact ground state wave function solution to (\ref{equation}) (upper curve) and closest 
approach using plane waves (lower curve). Notice the nodes at $\omega=0,\pi$. See text for details.}
\label{fig1}
\end{center}
\end{figure} 

Let us suppose now that want to use our approach using plane waves, which seems to be the most natural choice. 
In point of fact, we have just the replaced the Coulombian potential for the harmonic oscillator. However, as we shall see now, 
this approach fails quite dramatically.

Proceeding as previously, that is, substituting (\ref{spanned}) in (\ref{equationw}), multiplying by $\frac{1}{2\pi\sqrt{R_1 R_2}} e^{-ik\phi_1}e^{-il\phi_2}$ and integrating over $\{\phi_1,\phi_2\}$, we obtain

\begin{eqnarray} \label{set2}
\sum_{k=-\frac{N}{2}}^{\frac{N}{2}} \sum_{l=-\frac{N}{2}}^{\frac{N}{2}}  && \bigg[\bigg(\frac{m^2}{2R_1^2} + \frac{n^2}{2R_2^2} \bigg)\delta_{k,m}\delta_{l,n} \cr
&&+\frac{1}{2}\Omega^2\, \langle kl |  \big[R_1^2+R_2^2-2R_1R_2\cos(\phi_1-\phi_2)\big]  |mn\rangle \cr
&&-E\,\delta_{k,m}\delta_{l,n} \bigg]\,c_{k,l}\,=\,0,
\end{eqnarray}

\noindent which further simplifies into 

\begin{eqnarray} \label{set3}
\sum_{k=-\frac{N}{2}}^{\frac{N}{2}} \sum_{l=-\frac{N}{2}}^{\frac{N}{2}}  && \bigg[\bigg(\frac{m^2}{2R_1^2} + \frac{n^2}{2R_2^2} \bigg)\delta_{k,m}\delta_{l,n} \cr
&&-R_1R_2\Omega^2\, \langle kl | \cos(\phi_1-\phi_2) |mn\rangle \cr
&&-\Lambda_E\,\delta_{k,m}\delta_{l,n} \bigg]\,c_{k,l}\,=\,0,
\end{eqnarray}

\noindent with $\Lambda_E \equiv E\,-\,\frac{1}{2}\Omega^2(R_1^2+R_2^2)$. Matrix elements 
$\langle kl | \cos(\phi_1-\phi_2) |mn\rangle$ are different from zero for special values of the indexes. In any case, 
when we need to check the solution for the ground state, the maximum approach to the exact ground state wave function 
is far from being optimal. The only consistent fact is that the ensuing solution via basis truncation has real coefficients, which is tantamount as saying that it depends on $\phi_1-\phi_2$, as 
it is the case. The set of values for the ground state is given in Table II.

The corresponding wave function is compared with the exact one in Fig. 1. We can appreciate that 
no nodes are attained. The only way of obtaining these nodes by spanning the wave function in the basis of free particles 
in a quantum ring is when coefficients are such that sum of product plane-waves returns purely imaginary terms, sinus circular functions. 
However, this instance is not reached for some unknown reason.

\begin{table}[h]
\begin{center}
\begin{tabular}{|c||c|c|}
  \hline
  $k$ & $l$ & $c_{k,l}$ \\
  \hline
 -7 & 7 & 1.32216148E-007\\
 -6 & 6 &  4.21453413E-006\\
 -5 & 5  & 9.99675725E-005\\
 -4 & 4  & 0.00168284744\\
 -3 & 3  & 0.0188339041\\
 -2 & 2 &  0.127820814\\
 -1 & 1  & 0.460633757\\
 0 & 0 &  0.736370591\\
 1 & -1 &  0.460633757\\
 2 & -2 &  0.127820814\\
 3 & -3  & 0.0188339041\\
 4 & -4  & 0.00168284744\\
 5 & -5  & 9.99675725E-005\\
 6 & -6  & 4.21453413E-006\\
 7 & -7  & 1.32216148E-007\\
   \hline
\end{tabular}
\end{center}
\caption{Solution coefficients $c_{k,l}$ for the case $R_1 = 1$, $R_2=2$ for two particles interacting via Hooke's law. See text for details.}
\end{table}

Thus, having seen how well the truncation basis method works for electrons interacting via Coulomb repulsion as opposed to 
particles under Hooke's law, it is tantalizing to conclude that spanning the solution to the Schr\"odinger equation in the natural basis of the concomitant non-interacting system is not enough to ensure the correctness of that solution. However, if we compare the ground state wave function obtained via basis truncation {\it and} the exact one for the hypersphere when $R_1=R_2=R$, which is analytic \cite{Pierre2} as well, they have exactly the same behavior, with no nodes at either $\omega=0,\pi$. 

Therefore, we can appreciate an anomaly as far as matrix quantum mechanics is concerned when regarding systems interacting via Hooke's law. The plane wave approximation seems to be valid only for Coulomb interaction, but not for the harmonic oscillator 
{\bf unless} we go to a specific dimension (concentric hyperspheres), where the approach becomes exact.

\section{Conclusions}

We have presented two simple yet non-trivial quantum physics systems where the nature of the Hamiltonian defines whether the matrix formalism is correct or not. By definition, spanning the solution of the Schr\"odinger equation in a complete basis is an exact problem, regardless of the Hamiltonian involved. In the present contribution we provide an example of a system where the correctness of the formalism works well for a Coulomb interaction, whereas for a harmonic oscillator type it does net reach any satisfactory solution. This problem has interesting echoes not only in unveiling the details of the matrix formalism in quantum physics with more than one particle, but also in the fact that there exists an inconsistency which cannot be accounted for. Incidentally, the counterexample provided constitutes a new system not considered previously in the past. It is imperative to stress the fact that no errors due to truncation have to be considered because the approximation is extremely accurate.\newline

\section*{Acknowledgements}

J. Batle acknowledges fruitful discussions with J. Rossell\'o, Maria del Mar Batle and Regina Batle. J. Batle 
also appreciates fruitful discussions with Pierre-Francois Loos.

\end{document}